\documentclass{IEEEtran}
\usepackage{amssymb}
\usepackage{amsfonts}
\usepackage{amsmath}
\usepackage{algorithm}
\usepackage{graphicx,subfigure,cite,multicol,multirow,diagbox,booktabs,array}

\usepackage[dvips]{color}
\usepackage{float}
\usepackage{stfloats}
\UseRawInputEncoding

\ifCLASSINFOpdf
\else
\fi
\begin{document}
\title{Impact of Low-Resolution ADC on DOA Estimation Performance for Massive MIMO Receive Array}

\author{Baihua Shi, Nuo Chen, Xicheng Zhu, Yuwen Qian, Yijin Zhang, Feng Shu and Jiangzhou Wang,~\emph{Fellow},~\emph{IEEE}
\thanks{B. Shi , N. Chen, X. Zhu, Y. Qian and Y. Zhang are with the School of Electronic and Optical Engineering, Nanjing University of Science and Technology, Nanjing 210094, China.}
\thanks{F. Shu is with the School of Information and Communication Engineering, Hainan University, Haikou 570228, China. and also with the School of Electronic and Optical Engineering, Nanjing University of Science and Technology, Nanjing 210094, China. Email: {shufeng0101@163.com}.}
\thanks{J. Wang is with the School of Engineering and Digital Arts, University of Kent, Canterbury CT2 7NT, U.K. (e-mail: j.z.wang@kent.ac.uk).}
}
\maketitle

\begin{abstract}
    In this paper,  we present a new scenario of direction of arrival (DOA) estimation using massive multiple-input multiple-output (MIMO) receive array with low-resolution analog-to-digital convertors (ADCs), which  can strike a good balance between performance and circuit cost.  Based on the linear additive quantization noise model (AQNM),
    the effect of low-resolution ADCs on the methods, such as Root-MUSIC method, is analyzed. Also, the closed-form expression of Cramer-Rao lower bound (CRLB) is derived to evaluate the performance loss caused by the low-resolution ADCs. The simulation results show that the Root-MUSIC methods can achieve the corresponding CRLB. Furthermore, 2-3 bits are acceptable for most applications if the 1dB performance loss.
\end{abstract}


\IEEEpeerreviewmaketitle

\section{Introduction}
DOA estimation is important due to its diverse applications, including wireless communications, radar, navigation, and rescue and other emergency assistance devices~\cite{DOA,shu}. In recent application, such as internet of things (IoT),
angle of arrival (AOA) localization \cite{AOA},
massive multiple-input multiple-output (MIMO) and beyond so on \cite{Wang}, DOA estimation always plays an indispensable role.

Recently, as the massive MIMO becomes very popular, the DOA estimation using massive receive MIMO array emerges~\cite{shu}.
In \cite{Hu}, the deep-learning method was considered for the hybrid massive MIMO system with uniform circular array. However, the massive MIMO requires a large number of ADCs, which leads to a high circuit cost and energy consumption.
To solve this challenge, the electromagnetic (EM) lens antenna was considered for the DOA estimation in \cite{radioaccess}. The proposed methods could achieve a good performance.
Adopting low-resolution ADCs is another promising solution. Low-resolution ADCs have been used in many works~\cite{Zhang,oneXu,jinshi}. In \cite{jointtwc}, authors investigated the target detection and localization with low-resolution ADCs.
In \cite{Mengonebit}, a generalized sparse Bayesian learning algorithm was integrated into the 1-bit DOA estimation.
In \cite{onebit}, the authors proved that the MUSIC method can be straightforwardly applied without extra pre-processing and analysed the performance loss when the bit length is $1$.

To the best of our knowledge, although the DOA estimation for the one-bit ADCs has been investigated, no research work has been reported for the performance analysis of the ULA with low-resolution ADCs.
Thus, in this paper, we will derive the expression of performance loss which depends on the bit length of ADCs. By simulation, we will find the required minimum number of bits for ADCs when the performance loss is neglected.
It has the instructive significance on the practical applications.
The main contributions of this paper are summarized as follows:

\begin{enumerate}
    \item First, the system model of the DOA estimation for the low-resolution ADC structure is established by using the linear additive quantization noise model (AQNM). Then, based on this model, the effect of the low-resolution ADCs on the practical applications is analyzed. The Root-MUSIC is chosen as a representation of all related DOA estimation methods.
    \item  To evaluate the performance of the system, we derive the closed-form expression of the Cramer-Rao lower bound (CRLB) for the low-resolution ADC structure. Then, to concretize the performance loss caused by the low-resolutions, a performance loss factor is firstly defined by us. After Monte Carlo simulation, we find that ADCs with 2-3 bits can significantly reduce the circuit cost with a negligible performance loss, which is applied to many future applications.
\end{enumerate}

\emph{Notations:} Throughout the paper, matrices, vectors, and scalars are denoted by letters of bold upper case, bold lower case, and lower case, respectively. Signs $(\cdot)^T$, $(\cdot)^H$, $\mid\cdot\mid$ and $\parallel\cdot\parallel$ represent transpose, conjugate transpose, modulus and norm, respectively. $\textbf{I}_M$ denotes the $M\times M$ identity matrix. $\textbf{Tr}(\cdot)$ denotes matrix trace.
$\mathbb{E}[\cdot]$ represents the expectation. $\textbf{diag}(\cdot)$ denotes the diagonal operator. $arg(\cdot)$ means the argument of a complex number.

\section{System Model}
\begin{figure}
  \centering
  \includegraphics[width=0.35\textwidth]{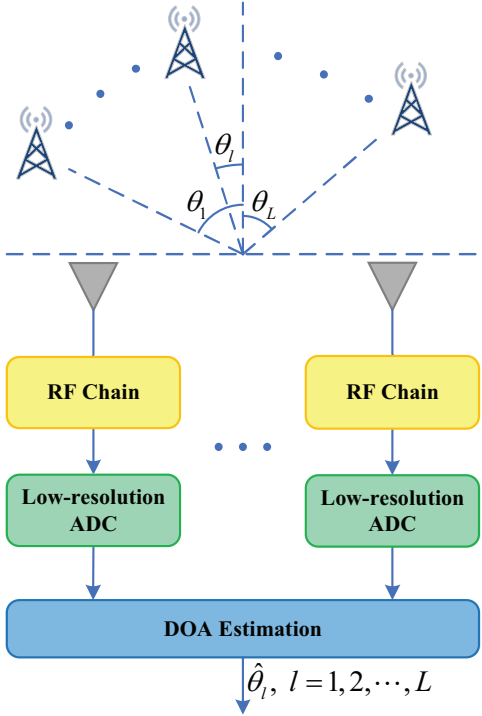}\\
  \caption{Architecture of receive array with low-resolution ADCs.}\label{sys_mod}
\end{figure}

As shown in Fig. \ref{sys_mod}, the signal will impinge on the uniformly-spaced linear array (ULA) which has $M$ antenna elements, and the different sensors will capture the same signals from far emitters with different time delays which depend on the DOAs. All signals are assumed as narrowband with the same carrier frequency $f_{c}$. Then, different from the traditional structure, the signal is digitized by low-resolution ADCs. The signals from far-field emitters are denoted by $s_{l}(t),l=1,2,\cdots,L$. Thus, a single emitter from the DOA $\theta$ can be modeled by
\begin{equation}\label{y_t_s}
\textbf{y}(t)=\textbf{a}(\theta)s(t)+\textbf{w}(t)
\end{equation}
where $\textbf{w}(t)\sim\mathcal{CN}(0,\sigma_{w}^{2}\textbf{I}_{M})$ is the additive white Gaussian noise (AWGN) vector and $\textbf{a}(\theta)$ is the so-called array manifold defined by
\begin{equation}\label{a_theta}
\textbf{a}(\theta)=[e^{j2{\pi}\psi_{\theta}(1)},e^{j2{\pi}\psi_{\theta}(2)},\cdots,e^{j2{\pi}\psi_{\theta}(M)}]^T,
\end{equation}
where $\psi_{\theta}(m)$ is the phase shift of signal at the baseband corresponding to the time delay from the source to antenna elements. $\psi_{\theta}(m),m=1,2,\cdots,M$, is given by
\begin{equation}\label{tao}
\psi_{\theta}(m)=\frac{d_{m}\sin\theta}{\lambda}, m=1,2,\cdots,M,
\end{equation}
where $\lambda$ is the wavelength of the carrier frequency.
$d_{m}$ is the distance from a common reference point to the $m$th antenna. In this letter, we choose the center of the array as the reference point. Thus, $d_m=\left(m-\frac{M}{2}\right)d$. $d$ is is half of the wavelength as usual, (i.e., $d = \lambda/2$).
Let consider a more common assumption that there are $L(L<M)$ emitters and their signals are uncorrelated.
Then, the received $M$ dimensional vector $\mathbf{y}$ can be expressed as
\begin{equation}\label{y_t}
\textbf{y}(t)=\sum\limits_{l=1}^{L}{\textbf{a}(\theta_{l})s_{l}(t)}+\textbf{w}(t)
\end{equation}
\begin{table}
\footnotesize
\centering
\caption{Distortion Factor $\beta$ for Different $b$-Bit ADCs}
\label{tab1}
\begin{tabular}{c|c|c|c|c|c}
\hline
$b$     & 1     & 2     & 3         & 4         & 5     \\
\hline
$\beta$ & 0.3634& 0.1175& 0.03454   &0.009497   &0.002499     \\
\hline
\end{tabular}
\end{table}
Now, we follow the general assumption that the all received signals $s_l(t)$ are independent Gaussian distributed and the quantization noise is the worst case of Gaussian distribution.
According to the widely used AQNM~\cite{dong},~\cite{arxiv}, which is able to convert the nonlinear quantization into the linear quantization gain with a additive quantization noise, the output signal vector of the low-resolution ADCs is given by
\begin{align}\label{y_q}
\textbf{y}_{q}(n)=\mathbb{Q}\{\textbf{y}(t)\}=\alpha\sum\limits_{l=1}^{L}{\textbf{a}(\theta_{l})s_{l}(t)}&+\alpha\textbf{w}(n)+\textbf{w}_{q}(n),\nonumber\\
&n=1,2,\cdots,N,
\end{align}
where $\mathbb{Q}\{\cdot\}$ is the quantization function, $N$ is the number of snapshots, $\alpha=1-\beta$ is the linear quantization gain,
where $\beta=\frac{\mathbb{E}[\|\textbf{y}-\textbf{y}_{q}\|^{2}]}{\mathbb{E}[\|\textbf{y}\|^{2}]}$ denotes the distortion factor caused by low-resolution ADCs \cite{dong},
and $\textbf{w}_{q}(n)$ is the quantization noise uncorrelated with $\textbf{y}$. We assume that the input of ADCs is Gaussian. Then, for the distortion-minimizing scalar non-uniform quantization,
the values of $\beta$ for $b\leq5$ are listed in Table \ref{tab1} and the other values of $\beta$ can be approximated by $\beta=\frac{\sqrt{3}\pi}{2}\cdot 2^{-2b},~b\geq 6.$ \cite{Zhang}.

For a fixed channel realization, the covariance matrix of $\textbf{w}_{q}(n)$ is expressed by
\begin{equation}\label{R_wq}
\textbf{R}_{\textbf{w}_q}=\alpha\beta \textbf{diag}(\sum\limits_{l=1}^{L}\sigma_{s,l}^{2}\textbf{a}(\theta_{l})\textbf{R}_{l}\textbf{a}(\theta_{l})^{H}+\sigma_{w}^{2}\textbf{I}_{M}),
\end{equation}
where $\sigma_{s,l}^{2}=\mathbb{E}[s_{l}(t)^{\ast}s_{l}(t)]$ is the power of the $l$th signal and $\textbf{R}_{l},l=1,2,\cdots,L$ is the $l$th normalized signal covariance matrix, so $\textbf{R}_{l}=\textbf{I}_{M},l=1,2,\cdots,L$ and (\ref{R_wq}) can be given by
\begin{align}\label{R_we}
\textbf{R}_{\textbf{w}_q}&=\alpha\beta \textbf{diag}\left(\sum\limits_{l=1}^{L}\sigma_{s,l}^{2}\textbf{a}(\theta_{l})\textbf{a}(\theta_{l})^{H}+\sigma_{w}^{2}\textbf{I}_{M}\right)\nonumber\\
&=\alpha\beta \textbf{diag}\left(\left(\sum\limits_{l=1}^{L}{\sigma_{s,l}^{2}+\sigma_{w}^{2}}\right)\textbf{I}_{M}\right).
\end{align}
For tractable analysis according to \cite{arxiv}, $\textbf{w}_{q}(n)$ can be modelled as $\textbf{w}_{q}(n)\sim\mathcal{CN}(0,\textbf{R}_{\textbf{w}_q})$.

\section{Analysis of  Root-MUSIC for Low-Resolution ADC Structure}
In this section, we analyze the impact of the low-resolution ADCs on the algorithms. As one of the most famous DOA methods, Root-MUSIC is chosen to be reviewed and analyzed.
The covariance matrix of the signal vector after quantization is given by
\begin{align}\label{R_yy}
\textbf{R}_{\textbf{y}_{q}\textbf{y}_{q}}&=\mathbb{E}[\textbf{y}_{q}\textbf{y}_{q}^{H}] \nonumber\\
&=\sum\limits_{l=1}^{L}{\alpha ^2\sigma_{s,l}^{2}\textbf{a}(\theta_{l})\textbf{a}(\theta_{l})^{H}}+\sigma_L^{2}\textbf{I}_{M},
\end{align}
where
\begin{equation}\label{sigma_L}
\sigma_L^{2}=\alpha ^2\sigma_{w}^{2}+\alpha\beta(\sum\limits_{l=1}^{L}{\sigma_{s,l}^2}+\sigma_{w}^{2})
\end{equation}
Let
\begin{equation}\label{R_svd}
\textbf{R}_{\textbf{y}_{q}\textbf{y}_{q}}=[\textbf{U}_{S}\ \textbf{U}_{N}]\Lambda[\textbf{U}_{S}\ \textbf{U}_{N}]^{H}
\end{equation}
be the eigenvalue decomposition of the (\ref{R_yy}) and
\begin{equation}\label{Lambda}
\Lambda=\textbf{diag}(\alpha^2\sigma_{s,1}^{2}+\sigma_L^{2},\cdots,\alpha^2\sigma_{s,l}^{2}+\sigma_L^{2},\sigma_L^{2},\cdots,\sigma_L^{2})
\end{equation}
where $M\times L$ matrix $\textbf{U}_{S}$ denotes the $L$ column vectors consisting of the singular vector corresponding to the largest singular values and the matrix $\textbf{U}_{N}$ contains the $(M-L)$ singular vectors corresponding to the $(M-L)$ smallest singular values.
Then, we can compute
\begin{equation}\label{pseudo}
S(\theta)=\|\textbf{U}_{N}^{H}\textbf{a}(\theta)\|_{2}^{-2}
\end{equation}
which will have peaks at the emitters' directions. The covariance matrix $\textbf{R}_{\textbf{y}_{q}\textbf{y}_{q}}$ is estimated from the available data by
\begin{equation}\label{R_yy_e}
\hat{\textbf{R}}_{\textbf{y}_{q}\textbf{y}_{q}}=\frac{1}{N}\sum\limits_{n=1}^{N}\textbf{y}_{q}(n)\textbf{y}_{q}(n)^{H}
\end{equation}

To obtain the emitter direction by maximizing (\ref{pseudo}), Root-MUSIC has low-complexity and near analytic solution without linear search. Root-MUSIC suggests to solve the following rooting polynomial
\begin{align}\label{f_root}
f_{root}(z)&=\mathbf{p}(z)^{H}\textbf{U}_{N}\textbf{U}_{N}^{H}\mathbf{p}(z)\nonumber\\
&=z^{M-1}\mathbf{p}^{T}(z^{-1})\textbf{U}_{N}\textbf{U}_{N}^{H}\mathbf{p}(z)=0,
\end{align}
where $\mathbf{p}(z)=[1,z,\cdots,z^{M-1}]^{T}$. The equation has $2(M-1)$ roots and $L$ pairs of roots that lie closest to the unit circle, where $z_{l}=e^{j2\pi \frac{d\sin\theta_{l}}{\lambda}},\ l=1,2,\cdots,L$ and $a(\theta_{l})=p(z_{l}),\ l=1,2,\cdots,L$. The DOA can be estimated by
\begin{equation}\label{theta_e}
\hat{\theta}_{l}=\arcsin\left( \frac{\lambda}{2\pi d}\arg z_{l}\right),~l=1,2,\cdots,L.
\end{equation}

Throughout the analysis above, we can conclude that the effect of low-resolution ADCs on the Root-MUSIC can be regarded as a reduction of the SNR. Thus, low-resolution ADCs will not affect the algorithm itself. In other words, computational complexity of the method is invariant for different quantization bits. Furthermore, this conclusion could be extended to other methods, such as ESPRIT.

\section{Derivation of CRLB}
The Cramer-Rao Lower bound (CRLB) for low-resolution ADC structure is derived in the following to evaluate the estimation performance of the array with $b$-bit ADCs.
Firstly, the Fisher Information Matrix (FIM) can be expressed by
\begin{equation}\label{F}
\textbf{F}_q=\textbf{Tr}\left\{ \textbf{R}_{\textbf{y}_{q}}^{-1} \frac{\partial \textbf{R}_{\textbf{y}_{q}}}{\partial \theta} \textbf{R}_{\textbf{y}_{q}}^{-1} \frac{\partial \textbf{R}_{\textbf{y}_{q}}}{\partial \theta} \right\},
\end{equation}
where $\textbf{R}_{\textbf{y}_{q}}$ can be given by
\begin{align}\label{R_yq}
\textbf{R}_{\textbf{y}_{q}}=\mathbb{E}[\textbf{y}_{q}\textbf{y}_{q}^{H}]
=\alpha^2\sigma_{s}^{2}\textbf{a}(\theta)\textbf{a}(\theta)^{H}+\sigma_{1}^{2}\textbf{I}_{M},
\end{align}
where, similar to the (\ref{sigma_L}),~$\sigma_{1}^{2}=\alpha ^2\sigma_{w}^{2}+\alpha\beta(\sigma_s^2+\sigma_{w}^{2})$. Given $N$ independent measurements, the CRLB is given by
\begin{equation}\label{CRLB}
\emph{CRLB}_b=\frac{1}{N}\textbf{F}_{q}^{-1}.
\end{equation}
Then, similar to \cite{DOA},
\begin{align}\label{F_e}
\textbf{F}_q&=2\gamma_{\textbf{y}_{q}}(2{\pi} / \lambda)^2 \cos^{2}\theta \overline{d^2}\frac{M\gamma_{\textbf{y}_{q}}}{M\gamma_{\textbf{y}_{q}}+1}\nonumber\\
&\approx2\gamma_{\textbf{y}_{q}}(2{\pi} / \lambda)^2 \cos^{2}\theta \overline{d^2}.
\end{align}
Thus, the CRLB can be derived by
\begin{equation}\label{CRLB_exp}
\emph{CRLB}_b \approx\frac{1}{2N \gamma_{\textbf{y}_{q}}(2{\pi} / \lambda)^2 \cos^{2}\theta \overline{d^2}},
\end{equation}
where
\begin{equation}\label{gamma_yq}
\gamma_{\textbf{y}_{q}}=\frac{\alpha\sigma_s^2}{\beta \sigma_s^2+\sigma_w^2}
\end{equation}
and
\begin{equation}\label{d_m}
\overline{d^2} = \sum\limits_{m=1}^{M}{d_{m}^{2}}.
\end{equation}

Now, let us define the performance loss factor in dB
\begin{equation}\label{R_CRLB}
\eta_{b}(\gamma)=10\log_{10}\frac{CRLB_{b}}{CRLB_{\infty}}
=10\log_{10}\frac{1+\beta\gamma}{\alpha} 
\end{equation}
where $\gamma=\sigma_s^2/\sigma_w^2$ is the input SNR of ADCs and $CRLB_{\infty}$ denotes the CRLB with high-resolution ADCs. From (\ref{R_CRLB}), it is obvious that the CRLB performance loss factor $\eta_{b}(\gamma)$ is a monotonically increasing function of $\gamma$ for a fixed $b$.


\section{Simulation and Discussion}
In this section, we present simulation results to find how many bits of ADCs to achieve an acceptable or omitted performance loss.
The performance loss factor is calculated by (\ref{R_CRLB}), and the RMSE is given by $RMSE=\sqrt{\frac{1}{N_t}\sum_{n_t=1}^{N_t}(\hat{\theta}_{n_t} - \theta)^2}$. All results are averaged over 8000 Monte Carlo realizations.
Simulation parameters are chosen as shown in Table \ref{tab2}.
\begin{table}
\footnotesize
\centering
\caption{Simulation Parameters}
\label{tab2}
\tabcolsep 40pt 
\begin{tabular}{c|c}
\hline
\hline
$\theta$     & $15^\circ$          \\
\hline
$M$ & 128     \\
\hline
$d$ & $\frac{\lambda}{2}$     \\
\hline
$N$ & 32     \\
\hline
$N_t$ & 8000\\
\hline
\hline
\end{tabular}
\end{table}

\begin{figure}[ht]
  \centering
  \includegraphics[width=0.45\textwidth]{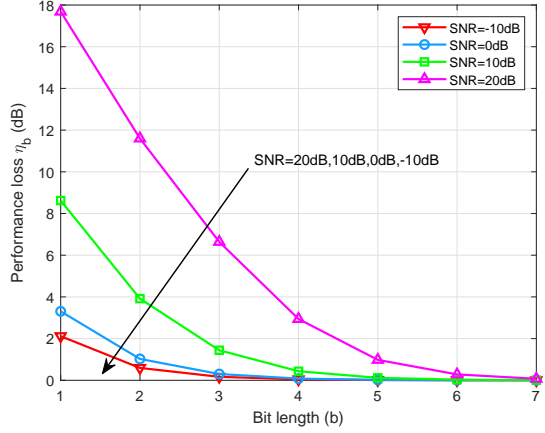}\\
  \caption{Performance loss factor versus bit length with $N=32$, $M=128$ for different SNR.}\label{CRLB_eta}
\end{figure}
Fig. \ref{CRLB_eta} illustrates the performance loss factor $\eta_b$ versus the bit length for the different SNR $\gamma$: $-20$dB, $-10$dB, $0$dB, $10$dB and $20$dB. It is clear that
the performance loss will be trivial, which means $\eta_b<1$dB, at the low and medium SNR when $b= 2$. As SNR decreases, the associated required minimum bit length reduced accordingly. Conversely, as SNR increases, the required bit length will increase gradually.
\begin{figure}[ht]
  \centering
  \includegraphics[width=0.45\textwidth]{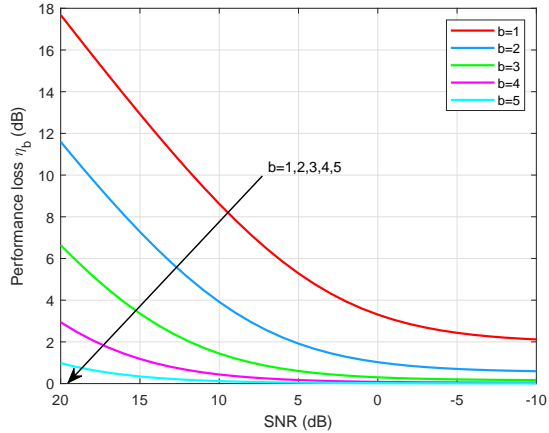}\\
  \caption{Performance loss factor versus the SNR with $N=32$, $M=128$ for different $b$.}\label{CRLB_eta_snr}
\end{figure}

Fig. \ref{CRLB_eta_snr} demonstrates the curves of performance loss factor versus  SNR with different $b$. From this figure, it can be seen that the performance loss factor $\eta_b$ increases as the SNR increases with a fixed bit length.
It is hard to accept that a performance loss with $b=1$ is higher than 2dB for all SNR. In addition, at low SNR ($\gamma<0dB$), 2-bit ADCs are satisfactory. $b=3$ is a better choice at medium SNR ($0<\gamma<10dB$). ADCs with 4-bits or 5-bits are more suitable for high SNR ($\gamma>10dB$). However, for most applications, the SNR is not high. Thus, 2-3 bits are sufficient. ADCs with more quantization bits could be adopted for ultra high precision requirements or applications at high SNR.


\begin{figure}[ht]
  \centering
  \includegraphics[width=0.45\textwidth]{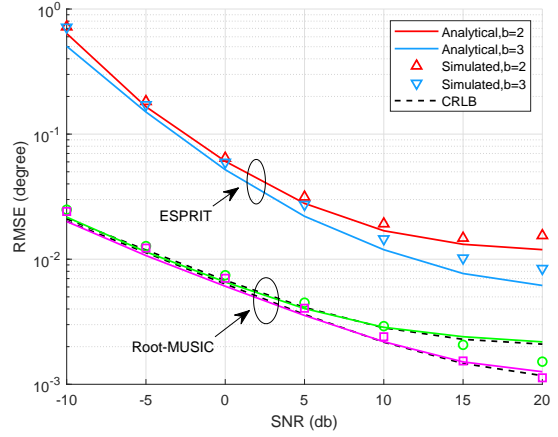}\\
  \caption{RMSE versus SNR for Root-MUSIC and ESPRIT.}\label{RMSE_snr}
\end{figure}

Fig. \ref{RMSE_snr} plots the RMSE versus SNR of the Root-MUSIC and ESPRIT for $b=2$ and $b=3$, with CRLB as a performance benchmark. Both the analytical and simulated results are presented. Herein, $\gamma$ ranges from $-10$dB to $20$dB. Observing Fig. \ref{RMSE_snr}, we find that all simulated results of the  Root-MUSIC and ESPRIT can achieve the associated analytical values at middle and high SNRs. As SNR increases, the gaps between the analytical and simulated curves become gradually large. This may be caused by the AQNM, because the AQNM is accurate enough at low and medium SNRs. In addition, the Root-MUSIC has better performance, which could achieve the corresponding CRLB.

%



\section{Conclusion}
In this paper, a low-cost framework, combining low-resolution ADCs and large-scale received MIMO, was proposed for DOA estimation to make a good balance between performance and circuit cost. The impact of low-resolution ADCs on the CRLB of DOA estimation was derived to show that the performance loss of variance of DOA estimate increases as SNR increases.
More importantly, the results have shown that 2-3 bits are sufficient for most applications. More bits should be adopted to achieve a better performance at medium and high SNRs.

\ifCLASSOPTIONcaptionsoff
  \newpage
\fi

\bibliographystyle{IEEEtran}
\bibliography{IEEE_cite}

\end{document}